\documentclass[twocolumn,showpacs,amsmath,amssymb]{revtex4}
\newcommand{\be}{\begin{eqnarray}}
\newcommand{\en}{\end{eqnarray}}
\newcommand{\ben}{\begin{eqnarray*}}
\newcommand{\enn}{\end{eqnarray*}}
\newcommand{\pa}{\partial}
\newcommand{\na}{\nabla}
\newcommand{\f}{\frac}

\newcommand{\bi}{\begin{itemize}}
\newcommand{\ei}{\end{itemize}}

\newcommand{\ld}{\lambda}

\newcommand{\va}{\varepsilon}

\newcommand{\R}{\Rightarrow}
\begin{document}
\title{{\Large{\it{\bf{On the MHD load and the MHD metage}}}}}
\author{Sagar Chakraborty}
\email{sagar@bose.res.in}
\affiliation{S.N. Bose National Centre for Basic Sciences, Saltlake, Kolkata 700098, India}
\author{Partha Guha}
\email{partha@bose.res.in}
\affiliation{S.N. Bose National Centre for Basic Sciences, Saltlake, Kolkata 700098, India}
\affiliation{Max Planck Institute for Mathematics in the Sciences,
Inselstrasse 22, D-04103 Leipzig, Germany}
\date{\today}
\begin{abstract}
In analogy with the load and the metage in hydrodynamics, we define magnetohydrodynamic load and magnetohydrodynamic metage in the case of magnetofluids.
They can be used to write the magnetic field in MHD in Clebsch's form.
We show how these two concepts can be utilised to derive the magnetic analogue of the Ertel's theorem and also, how in the presence of non-trivial topology of the magnetic field in the magnetofluid one may associate the linking number of the magnetic field lines with the invariant MHD loads.
The paper illustrates that the symmetry translation of the MHD metage in the corresponding label space generates the conservation of cross helicity.
\end{abstract}
\pacs{47.65.-d, 52.30.Cv}
\maketitle
\section{Introduction}
\indent In a pioneering paper\cite{Bell}, Lynden-Bell and Katz introduced the concept of load and metage in the hydrodynamical flows to show that the content of Kelvin's circulation theorem is contained in the conservation of a particular function of load; if that very function is same for two flows then the flows are isocirculational.
In the case of MHD flows, one can define similar analogous quantities that, as we shall see in this paper, can prove to be quite significant both physically and mathematically.
\\
\indent A magnetofluid (also termed hydromagnetic fluid or MHD fluid) of infinite conductivity but possibly compressible and viscous is quite an interesting class of fluids in which the flow velocity ($\vec{v}$) and the magnetic field ($\vec{B}$) interact actively; in general, $\vec{B}$ cannot be considered to be a passively advected quantity.
Anyway, owing to the Alfv$\acute{\textrm{e}}$n's theorem, $\vec{B}$ remains frozen in the magnetofluid; as we shall see, this property helps to design three active scalars that remain conserved on the fluid particles allowing themselves to be used as the particle labels which by definition, although vary continuously throughout the fluid, remain fixed on the fluid particles and hence serve as possible coordinates of the label space in the Lagrangian description of the flow.
With the comment that these scalars may become passive scalars in certain well-studied cases {\it e.g.,} in kinematic dynamo problem\cite{Choudhuri} wherein one basically studies the instability around $\vec{B}=0$ state of MHD because there the magnetic field itself is treated to be passively advected, let us now look for the aforementioned advected scalars.
\section{MHD load and MHD metage}
\indent We take an isolated volume of magnetofluid of negligible resistivity.
$\vec{B}$ will be penetrating it all throughout.
Due to solenoidal nature, some magnetic field lines will be entirely inside the magnetofluid and some will penetrate out of the boundaries and complete the loop outside the magnetofluid.
Lets assume that the later type of field lines begin at some the initial point on the surface which we term as ``entry'' to leave the magnetofluid at some corresponding final points on the boundary that we call ``exit''.
Having taken such a line, we construct a magnetic flux tube of small cross-section around the line which serves as a sort of axis to the tube.
Obviously, for the magnetic field loops lying entirely within the magnetofluid, the ``entry'' and ``exit'' collapse onto the same point and hence one should keep in mind that in the calculation that follows one would need to introduce a ``cut'' somewhere on the flux tube defined about such a loop.
We also assume for the time being that the magnetic field lines are not knotted.
\\
\indent Let $\Delta S(l)$ be the cross-section of the tube at a distance $l$ from the entry and at that point let the mass of the infinitesimal disk of fluid of thickness $dl$ be $dm$. So, mass of the tube (up to the first order in $\Delta S(l)$) is
\be
\Delta m=\int_\textrm{entry}^\textrm{exit}dm=\int_\textrm{entry}^\textrm{exit}\rho(l)\Delta S(l)dl
\label{mass}
\en
Again, if $\Delta\phi$ is the strength of the flux tube then similarly one has
\be
\Delta \phi=B(l)\Delta S(l)
\label{flux}
\en
Using the relation (\ref{flux}) in the equation (\ref{mass}) and taking the limit as the cross section of the flux tube going to zero, one arrives at
\be
\lambda\equiv\f{dm}{d\phi}=\int_\textrm{entry}^\textrm{exit}\f{\rho(l)}{B(l)}dl
\label{MHDL}
\en
where $\lambda$ may be called ``magnetohydrodynamic load'' in analogy with the ``load'' in hydrodynamics.
Now, we give this scalar $\lambda$ the status of the scalar field by defining $\lambda$ at a point as the $\lambda$ of the field line passing through that point.
As, $\lambda$ is constant along the magnetic field line, so it must satisfy the equation:
\be
\vec{B}.\vec{\nabla}\Lambda=0
\label{Lambda}
\en
{\it i.e.}, $\Lambda=\tilde{\lambda}$ is a solution of the relation (\ref{Lambda}); we have
\be
\vec{B}.\vec{\nabla}\lambda&=&0
\label{lambda1}
\en
and as $\lambda$ remains constant with the motion of the field line due to the frozen condition of the magnetic field, we can write the following mathematical form for the condition:
\be
\f{D\lambda}{Dt}&=&0
\label{lambda2}
\en
Again, let $\Lambda=\tilde{\lambda}$ be another solution of the relation (\ref{Lambda}) such that
\be
\f{\pa\tilde{\lambda}}{\pa\lambda}=0
\label{0}
\en
So, one has like the equation (\ref{lambda1}), a similar equation for $\tilde{\lambda}$ that we shall call ``conjugate magnetohydrodynamic load'' due to the reasons which will become obvious in due course.
$\tilde{\lambda}$ naturally satisfies (by definition) the following equation:
\be
\vec{B}.\vec{\nabla}\tilde{\lambda}=0
\label{lambdat1}
\en
From the equations (\ref{lambda1}) and (\ref{lambdat1}) we can write
\be
\vec{B}=C(\vec{\nabla}\lambda\times\vec{\nabla}\tilde{\lambda})
\label{B1}
\en
because $\vec{B}$ is orthogonal to both $\vec{\nabla}\lambda$ and $\vec{\nabla}\tilde{\lambda}$.
Here, due to the solenoidal nature of the magnetic field, one may conclude taking divergence of the relation (\ref{B1}) that
\be
C=C(\lambda,\tilde{\lambda})
\en
Redefining, $\tilde{\lambda}$ as $\int_0^{\tilde{\lambda}}C(\lambda,\tilde{\lambda})d\tilde{\lambda}$ we can do away with $C$ to get ultimately from the relation (\ref{B1}):
\be
\vec{B}=\vec{\nabla}\lambda\times\vec{\nabla}\tilde{\lambda}
\label{B2}
\en
which is the form of magnetic field associated with Clebsch parametrised magnetic vector potential.
Thus, now $\vec{B}$ automatically satisfies the relation $\vec{\na}.\vec{B}=0$.
A very interesting geometrical interpretation for the relation (\ref{B2}) is that the part of the field line inside the magnetofluid is basically the curve formed due to the intersection of the surface of constant $\lambda$ with the surface of constant $\tilde{\lambda}$.
Thus the specification of the families of the such surfaces is an alternative way of specifying the $\vec{B}$ in the magnetofluid; $\lambda$ and $\tilde{\lambda}$ though do not vary along the field lines, distinguish between the field lines.
\\
\indent A third scalar is now needed to complete the set of coordinates of the label space.
Let us refer the equation (\ref{MHDL}) again and change the upper limit from ``exit'' to some floating point (which we shall call ``float''), on the narrow tube about the field line, at a distance $l$ from the ``entry''.
Using the concepts of conservations of mass and magnetic strength, define the following Lagrangian invariant scalar which we define as ``magnetohydrodynamic metage'' $\mu$:
\be
\mu\equiv\f{dm_f}{d\phi}=\int_\textrm{entry}^\textrm{float}\f{\rho(l)}{B(l)}dl
\label{MHDM}
\en
where the subscript ``$f$'' on $m$ refers to the fact that one is interested only in the mass of the tube up to the length ending at the floating point.
Due to the frozen nature of the magnetic field line, the floating point moves with the fluid and this makes $\mu$ a candidate for labeling the fluid elements.
One, thus, has
\be
\f{D\mu}{Dt}=0
\label{mu2}
\en
Also, from the relation (\ref{MHDM}) it follows that
\be
\vec{B}.\vec{\nabla}\mu=\rho
\label{mu1}
\en
From the equations (\ref{lambda1}), (\ref{lambdat1}), (\ref{B2}) and (\ref{mu1}), one finds that the Jacobian $J$ is
\be
J\equiv\f{\pa(\lambda,\tilde{\lambda},\mu)}{\pa(x,y,z)}=(\vec{\nabla}\lambda\times\vec{\nabla}\tilde{\lambda}).\vec{\nabla}\mu=\rho
\label{J}
\en
Now this value of $J$ confirms that $\lambda$, $\tilde{\lambda}$ and $\mu$ are the independent variables in the Lagrangian description and we can think of a label space with coordinates ($\lambda$,$\tilde{\lambda}$,$\mu$).
The coordinate system ($\lambda$,$\tilde{\lambda}$,$\mu$) may be treated as curvilinear coordinate system in the usual $(x,y,z)$ space.
One can choose $\vec{\na}\lambda$, $\vec{\na}\tilde{\lambda}$ and $\vec{\na}\mu$ as basis vectors in this curvilinear coordinate system.
As $\tilde{\lambda}$ is a label so it must satisfy the following equation:
\be
\f{D\tilde{\lambda}}{Dt}=0
\label{lambdat2}
\en
So just like $\lambda$, $\tilde{\lambda}$ is a Clebsch type co-moving scalar field. Also, from the relations (\ref{lambda2}), (\ref{lambdat2}) and (\ref{B2}) the usual induction equation written below must follow.
\be
\f{\pa \vec{B}}{\pa t}=\vec{\na}\times(\vec{v}\times\vec{B})
\label{inductionequation}
\en
$\vec{v}$ being the velocity field of the MHD flow.
\section{$\lambda$ and magnetic Ertel's theorem}
\indent One can always define a vector potential $\vec{A}$ for $\vec{B}$.
Although $\vec{A}$ is always non-unique in the sense that gradient of any scalar field may always be added to it without changing the $\vec{B}$ and hence one would notice that in the arguments that follow it hardly matters which $\vec{A}$ one is choosing.
Anyway, w.r.t. the basis vectors of the coordinate system ($\lambda$,$\tilde{\lambda}$,$\mu$), $\vec{A}$ may be written as
\be
\vec{A}=A_1\vec{\na}\ld_1+A_2\vec{\na}\ld_2+A_3\vec{\na}\ld_3
\label{A}
\en
For convenience, we have renamed ($\lambda$,$\tilde{\lambda}$,$\mu$) as $(\ld_1,\ld_2,\ld_3)$.
Lets define
\be
\vec{{B}}_p\equiv\vec{\na}_{\ld}\times\vec{A}
\label{Bt}
\en
where, $\vec{\na}_{\ld}\equiv\left({\pa}/{\pa \ld_1},{\pa}/{\pa \ld_2},{\pa}/{\pa \ld_3}\right)$.
Seemingly, the relation (\ref{Bt}) suggests that $\vec{{B}}_p$ is the magnetic field measured in the curvilinear coordinate system ($\lambda$,$\tilde{\lambda}$,$\mu$).
We shall see (equation (\ref{relation})) that each component of $\vec{B}_p$ is a conserved quantity which may be interpreted as a potential magnetic field $B_{\ld}$ w.r.t some Lagrangian invariant scalar.
\\
\indent Potential magnetic field $B_{\ld}$ w.r.t any Lagrangian invariant scalar, say $\lambda$ for example, may defined as
\be
B_{\ld}\equiv\left(\f{\vec{B}}{\rho}.\vec{\na}\right)\ld
\label{pmf}
\en
Due to the frozen nature of the magnetic field which is the result of the following equation valid in the MHD of infinite conductivity:
\be
\f{D}{Dt}\left(\f{\vec{B}}{\rho}\right)=\left(\f{\vec{B}}{\rho}.\vec{\na}\right)\vec{v}
\label{frozen}
\en
As the infinitesimal displacement $\vec{\delta x}$ between two moving fluid particles follow equation similar to the equation (\ref{frozen}), we have
\be
\f{d\vec{\delta x}}{dt}=(\vec{\delta x}.\vec{\na})\vec{v}
\label{element}
\en
Due to the relation (\ref{lambda2}), one can easily put the equations (\ref{frozen}) and (\ref{element}) to good use and can churn out in the process the following equation:
\be
\f{D}{Dt}\left\{\left(\f{\vec{B}}{\rho}.\vec{\na}\right)\ld\right\}=0
\label{2}
\en
that may be simply interpreted as the conservation of $B_{\ld}$ ({\it i.e.} potential magnetic w.r.t.MHD load) on fluid particles; obviously this is the analogue of Ertel's theorem\cite{Salmon} in MHD.
\\
\indent The vector potential $\vec{A}$ (see relation (\ref{A})), may also be written as
\be
\vec{A}=A'_1\hat{i}+A'_2\hat{j}+A'_3\hat{k}
\label{A1}
\en
The transformation rule for the components of $\vec{A}$ in two different basis vectors (refer to relations (\ref{A}) and (\ref{A1}) is:
\be
A'_i=\f{\pa\ld_j}{\pa x_i}A_j
\label{trule}
\en
This rule (\ref{trule}) and the relation (\ref{J}) allows one to establish that
\be
(\vec{\na}_{\ld}\times\vec{A})_i={B}_{\ld_i}
\label{relation}
\en
This is what we wanted to arrive at.
Comparing the relations (\ref{Bt}) and (\ref{relation}) we can see that the magnetic field measured in the curvilinear coordinate system ($\lambda$,$\tilde{\lambda}$,$\mu$) is a conserved quantity.
\section{MHD loads and linking number}
\indent A useful function of $\lambda$ may be defined in a steady MHD flow.
Steady, by definition, means the explicit independence of the physically measurable quantities from time; hence $\rho$ and $\vec{B}$ do not depend on time in the case of steady MHD flow.
The definition of $\lambda$ (equation (\ref{MHDL})) would then suggest that the magnetohydrodynamic load as well is not dependent on time.
Lets define the function
\be
\Phi(\lambda)\equiv\int_{\lambda}\vec{B}.d\vec{S}
\label{Phi}
\en
that basically gives the strength of the flux through a large magnetic field tube surfaced by a locus of points having constant $\lambda$; for convenience of discussion we henceforth shall mean such a tube when we shall be using the phrase ``tube of load $\lambda$'' and shall denote the mass within it by $M(\lambda)$.
From the definitions (\ref{MHDL}) and (\ref{Phi}), we can very well write
\be
\f{dM}{d\lambda}=\lambda\f{d\Phi}{d\lambda}
\label{dMdl}
\en
If $\lambda=\lambda_0$ is the MHD load at the center line of the tube of load $\lambda$, one may note that
\be
M(\lambda_0)=0;\phantom{xxx}\Phi(\lambda_0)=0
\label{tworel1}
\en
Hence, from the equation (\ref{dMdl}) one notices that $\Phi(\lambda)$ and $M(\lambda)$ and MHD load can be related to each other by dint of the following relations:
\be
M(\lambda)&=&\int^{\lambda}_{\lambda_{0}}\lambda'\f{d\Phi(\lambda')}{d\lambda'}{d\lambda'}=\int_0^{\Phi(\lambda)}\lambda(\Phi')d\Phi'
\label{tworel2}\\
\Phi(\lambda)&=&\int^{\lambda}_{\lambda_{0}}\f{1}{\lambda'}\f{dM(\lambda')}{d\lambda'}{d\lambda'}=\int_0^{M(\lambda)}\f{dM'}{\lambda(M')}
\label{tworel3}
\en
One of the interesting mathematical applications of defining $\Phi(\lambda)$ is that through it the concept of MHD load $\lambda$ can go on to express magnetic helicity in its own terms as should be obvious from what follows.
\\
\indent Consider the case that the magnetic field lines spiral around toroids which lie entirely within the magnetofluid.
Such an arrangement would inherently introduce magnetic helicity in the flow.
This should be evident from the fact that one can build such a structure by ``Dehn surgery'', {\it i.e.} by cutting a closed magnetic flux tube at a section, twisting the free ends through a relative angle that is integral multiple of $2\pi$, and reconnecting; for simplicity's sake one may suppose that the resulting twist is uniformly distributed round the tube.
Obviously, two fluxes of magnetic lines of force are associated with each toroid: one being the flux threading the meridional cross-section of the toroid and the other being the one which penetrates the inner boundary of the equatorial section of the toroid.
Let the strengths of the fluxes be $\Phi_m$ and $\Phi_e$ respectively. In this non-trivial topology, one can define two invariant MHD loads to characterise the toroid as
\be
\lambda_m\equiv\f{dM}{d\Phi_m};\phantom{xxx}\lambda_e\equiv\f{dM}{d\Phi_e}
\label{tworel4}
\en
where $M$ is the mass inside the torus that provides the magnetic field lines a compact support $V$ on whose boundary $\pa V$, $\hat{n}.\vec{B}=0$.
\\
\indent Now, a conserved pseudo-scalar quantity {\it viz.}, magnetic helicity\cite{Moreau,Moffatt,Ranada} is given by:
\be
H\equiv\int_V\vec{A}.\vec{B}dV
\label{H}
\en
The integrand here measures how much $\vec{A}$ rotates about itself times its modulus; thereby making the name magnetic helicity apt because it gauges the relative curling and braiding of the lines of $\vec{A}$ and $\vec{B}$.
Thus, $H$ gives an intuitive measure as to what degree the field lines resemble helices.
The most important property of $H$ which is of our concern is that it takes non-vanishing values only if the topology of the integral curves of magnetic field is non-trivial.
Obviously in the definition (\ref{H}), $\vec{A}$ is not unique, for, a term $\vec{\na}\theta$, $\theta$ being a scalar field can always be added to it keeping $\vec{B}$ unchanged.
Let us ponder over this non-uniqueness of the vector potential\cite{Berger}.
For smooth discussion's sake, we assume for the time being that the volume is simply connected.
Suppose $A_i\rightarrow A_i+\pa_i\theta$, then from the definition (\ref{H}) of magnetic helicity we can find the change $\delta{H}$ in $H$ to be:
\be
\delta H=\int_V\vec{\na}\theta.\vec{B}d^3x=\oint_{\pa{V}}\theta\vec{B}.\hat{n}d^2x
\label{dH}
\en
where $\hat{n}$ is the unit vector perpendicular to the infinitesimal surface element $d^2x$ and we have used the solenoidal nature of magnetic field and Gauss divergence theorem.
The relation (\ref{dH}) amounts to saying that the magnetic helicity will be gauge invariant in case the surface $\pa V$ bounding $V$ is a magnetic surface {\it i.e.}, $\vec{B}.\hat{n}=0$ on $\pa V$.
This condition for gauge invariance is rather strong because if $\vec{B}.\hat{n}\ne 0$ on $\pa V$ then one cannot seek refuge in Coulomb gauge for it is too loosely defined inside $V$ with no information about the outside field whatsoever.
More starkly, it means that different solenoidal vector potentials inside $V$ can correspond to Coulomb potentials of fields which have different structures outside $V$.
Now if we relax the condition that $V$ is simply connected, then the line integrals of $\vec{A}$ about the `holes' in the possibly multiply connected region have to be specified in order to have gauge-invariant magnetic helicity within $\pa V$ on which $\vec{B}.\hat{n}=0$.
\\
\indent It is well-known\cite{Ricca} that if the flux strength in the tube before the `Dehn surgery' was $\Phi$ and the number of twists required to get the toroidal configuration mentioned above be $n$ ({\it i.e.}, the linking number of any pair of the magnetic field lines is $n$) then
\be
H=n\Phi^2
\label{n}
\en
But since in the configuration under study, one may use relations (\ref{tworel4}) to rewrite the definition (\ref{H}) as:
\be
H=\int_0^M{\Phi_e(M')}\left[\f{d}{dM'}\Phi_m(M')\right]dM'
\label{H2}
\en
It is of interest to note that using the relations (\ref{tworel4}), (\ref{n}) and (\ref{H2}), one may express the linking number in terms of the invariant MHD loads as in the following relation:
\be
n=\f{1}{\Phi^2}\int_{M'=0}^M\int_{M''=0}^{M'}\f{dM''dM'}{\lambda_e(M'')\lambda_m(M')}
\label{nl}
\en
This evidently provides a physical and topological significance for the MHD loads.
\section{$\mu$ and cross helicity}
\indent Like the topological conserved quantity magnetic helicity, in the non-viscous MHD with infinite conductivity one has yet another conserved quantity {\it viz.}, cross helicity defined as follows:
\be
H_c\equiv\int_V\vec{v}.\vec{B}dV
\label{Hc}
\en
It measures the degree of mutual knottedness\cite{Moffatt} of the vorticity field and the magnetic field; this remains constant even though the vortex lines are no longer frozen in the fluid.
In this section, we shall prove that an appropriate symmetry displacement associated with the infinitesimal change in the MHD metage $\mu$ generates cross helicity conservation.
A similar analysis\cite{Yahalom} with hydromagnetic load and metage yields conservation of helicity.
\\
\indent The equation for an ideal non-dissipative MHD flow is:
\be
\f{\pa \vec{v}}{\pa t}+(\vec{v}.\vec{\na})\vec{v}=-\f{\vec{\na}P}{\rho}-\f{\vec{B}\times(\vec{\na}\times\vec{B})}{\mu_0\rho}-\vec{\na}\psi
\label{MHDeqn}
\en
where $\psi$ is the external potential and $\mu_0$ is the permeability of the medium.
If $s$ is the specific entropy and $\varepsilon(\rho,s)$ is the specific internal energy, one can write the following Lagrangian corresponding to the equation (\ref{MHDeqn}):
\be
L=\int_V\left[\f{1}{2}v^2-\varepsilon(\rho,s)-\psi-\f{B^2}{2\mu_0\rho}\right]\rho d^3x
\label{L}
\en
The pressure $P$ is related to $\varepsilon(\rho,s)$ via the thermodynamical relation:
\be
P=\rho^2\left(\f{\pa\varepsilon}{\pa\rho}\right)_s
\label{P}
\en
The action is defined as:
\be
A\equiv\int_{t_1}^{t_2} Ldt
\label{A}
\en
Following the notations developed in the section (III), as $\vec{x}$ is the position vector of a particle labelled by $\vec{\lambda}=(\lambda_1,\lambda_2,\lambda_3)$, the trajectories are given by:
\be
\vec{x}=\vec{X}(\vec{\lambda},t)
\en
where $\vec{X}$ is a vectorial function that induces a mapping that if assumed one-to-one, can have an inverse.
We would distinguish between the variations $\Delta$ incurred on a given label ({\it i.e.}, at a point in the label space) and the variations $\delta$ at a fixed point in the real space.
If infinitesimal changes $\Delta\vec{X}$ are made in the trajectories, for any quantity $F$ one has the following relation\cite{Bretherton}:
\be
\delta F=\Delta F-\Delta\vec{X}.\vec{\na}F
\label{F}
\en
Now, we have the continuity equation and the induction equation as follows respectively:
\be
\f{D\rho}{Dt}&\equiv&\f{\pa \rho}{\pa t}+(\vec{v}.\vec{\na})\rho=-\rho(\vec{\na}.\vec{v})\label{continuity}\\
\f{D\vec{B}}{Dt}&\equiv&\f{\pa \vec{B}}{\pa t}+(\vec{v}.\vec{\na})\vec{B}=(\vec{B}.\vec{\na})\vec{v}\label{B}-\vec{B}(\vec{\na}.\vec{v})
\en
They respectively suggest following variations inducing changes in $\rho$ and $\vec{B}$:
\be
\Delta\rho&=&-\rho(\vec{\na}.\Delta\vec{X})\label{rho}\\
\Delta B_i&=&B_j\pa_j\Delta X_i\label{B}-B_i\pa_j\Delta X_j
\en
Similarly, as is generally assumed if the specific entropy $s$ of a fluid particle is constant, then
\be
\Delta s=0
\label{s}
\en
Also, the corresponding variation of $\vec{v}$ is:
\be
\Delta{v_i}=\f{D}{Dt}\Delta X_i
\label{v}
\en
Equipped with these relations, we proceed to apply variation on the action (\ref{A}) rewritten explicitly below
\be
A=\int_{t_1}^{t_2}dt\int_Vd^3x\left[\left\{\f{1}{2}v^2-\varepsilon(\rho,s)-\psi\right\}\rho-\f{B^2}{2\mu_0}\right]
\label{A2}
\en
We shall consider that suitably differentiable variations in the particle trajectories vanish for $t$ outside $[t_1,t_2]$ and $\Delta\vec{X}.\hat{n}=0$ on the boundary of $V$.
Therefore, we have starting from the expression (\ref{A2}):
\begin{widetext}
\be
\delta{A}&=&\int_{t_1}^{t_2}dt\int_Vd^3x\left[\left\{\f{v^2}{2}-\va-\psi\right\}\delta\rho+\rho\left\{\vec{v}.\delta\vec{v}-\delta\va-\delta\psi\right\}-\f{\vec{B}.\delta\vec{B}}{\mu_0}\right]\label{dA1}\\
\R\delta{A}&=&\int_{t_1}^{t_2}dt\int_Vd^3x\left[\left\{\f{v^2}{2}-\va-\psi\right\}(\Delta\rho-\Delta\vec{X}.\vec{\na}\rho)+\rho\left\{\vec{v}.\Delta\vec{v}-\vec{v}(\Delta\vec{X}.\vec{\na})\vec{v}-\Delta\va+\Delta\vec{X}.\vec{\na}\va-\Delta\psi+\Delta\vec{X}.\vec{\na}\psi\right\}\right.\nonumber\\
&{}&\left.-\f{1}{\mu_0}\left\{\vec{B}.\Delta\vec{B}-\vec{B}.(\Delta\vec{X}.\vec{\na})\vec{B}\right\}\right]\label{dA2}\\
\R\delta{A}&=&\int_{t_1}^{t_2}dt\int_Vd^3x\left[\left\{\f{v^2}{2}-\f{\pa}{\pa\rho}(\rho\va)-\psi\right\}\Delta\rho+\rho\vec{v}.\delta\vec{v}-\rho\f{\pa\va}{\pa s}\Delta{s}-\rho\vec{\na}\psi.\Delta\vec{X}-\Delta\vec{X}.\vec{\na}\left\{\f{\rho v^2}{2}-\rho\va-\rho\psi\right\}\right.\nonumber\\
&{}&\left.-\f{1}{\mu_0}\left\{B_i(B_j\pa_j)\Delta X_i-B_i(B_i\pa_j\Delta X_j)-\Delta\vec{X}.\vec{\na}\left(\f{B^2}{2}\right)\right\}\right]\label{dA3}\\
\R\delta{A}&=&\int_{t_1}^{t_2}dt\int_Vd^3x\left[\left(\vec{\na}\left\{\f{\rho v^2}{2}-\rho \f{\pa}{\pa\rho}(\rho\va)-\rho \psi\right\}-\f{D}{Dt}(\rho\vec{v})-\rho\vec{\na}\psi-\vec{\na}\left\{\f{\rho v^2}{2}-\rho\va-\rho\psi\right\}\right).\Delta\vec{X}\right.\nonumber\\
&{}&\left.-\f{1}{\mu_0}\left\{-\pa_j(B_iB_j)+\pa_iB^2-\f{1}{2}\pa_iB^2\right\}\Delta X_i\right]+\left[\int_Vd^3x\rho\vec{v}.\Delta\vec{X}\right]_{t_1}^{t_2}\label{dA4}\\
\R\delta{A}&=&\left[\int_Vd^3\lambda\vec{v}.\Delta\vec{X}\right]_{t_1}^{t_2}-\int_{t_1}^{t_2}dt\int_Vd^3x\left[\f{D}{Dt}(\rho\vec{v})+\vec{\na}P+\rho\vec{\na}\psi-\f{(\vec{B}.\vec{\na})\vec{B}}{\mu_0}+\vec{\na}\left(\f{B^2}{2\mu_0}\right)\right].\Delta\vec{X}\label{dA5}
\en
\end{widetext}
We pause here a bit and explain the steps involved in the above calculation.
To reach from (\ref{dA1}) to (\ref{dA2}), we have used (\ref{F}); and (\ref{dA2}) in turn used (\ref{B}) to yield (\ref{dA3}).
In arriving at (\ref{dA4}), relations (\ref{rho}), (\ref{s}) and (\ref{v}) have been taken into account followed by integration by parts of the appropriate terms.
The final step, relation (\ref{dA5}) follows from (\ref{dA4}) by simple rearrangement and by using the relation (\ref{P}).
At this point, we argue that the variation is a symmetry displacement ({\it i.e.}, a displacement that makes $\delta{A}$ zero) and allow the equations (\ref{MHDeqn}) and (\ref{continuity}) to hold during the process of variation.
Hence, the equation (\ref{dA5}) yields:
\be
\int_V(\vec{v}.\Delta\vec{X})d^3\lambda=\textrm{constant}
\label{constant}
\en
\indent Simply because one is always free to choose a new set of variables $(\lambda'_1,\lambda'_2,\lambda'_3)$ such that
\be
\f{\pa(\vec{\lambda'})}{\pa(\vec{\lambda})}\equiv\f{\pa(\lambda'_1,\lambda'_2,\lambda'_3)}{\pa(\lambda_1,\lambda_2,\lambda_3)}=1,
\en
the $\vec{\lambda}$ chosen to satisfy equation (\ref{J}) is not unique.
Hence, if the new set of variables does not modify the domain of integration, the value of the Lagrangian in the definition (\ref{L}) remains the same.
It is nothing but the fascinating concept of relabeling symmetry\cite{Salmon}.
\\
As for the magnetic field lines lying entirely within the hydromagnetic fluid are always closed, we can translate MHD metage $\mu$ without affecting the Lagrangian; this means that the appropriate symmetry displacement is:
\be
\Delta\vec{X}=-\f{\pa\vec{X}}{\pa\mu}\Delta\mu
\en
where, $\vec{x}=\vec{X}(\lambda,\tilde{\lambda},\mu)$ is the transformation whose Jacobian is given by the inverse of the Jacobian given in the relation (\ref{J}).
Thus, in the additional light of the fact that varying Lagrangian of the equation (\ref{L}) by a constant $\Delta\mu$ keeps $L$ unaffected, the relation (\ref{constant}) becomes:
\be
\int_V\vec{v}.\f{\pa\vec{X}}{\pa\mu}Jd^3x&=&\textrm{constant}\\
\R\int_V\vec{v}.\left(\rho\f{\pa\vec{X}}{\pa\mu}\right)d^3x&=&\textrm{constant}
\label{dot1}
\en
But the relation (\ref{mu1}) suggests\cite{Kuznetsov}
\be
\rho\f{\pa\vec{X}}{\pa\mu}=\vec{B}
\label{dot2}
\en
and therefore putting (\ref{dot2}) in (\ref{dot1}), we arrive at the result:
\be
H_c=\int_V\vec{v}.\vec{B}dV=\textrm{constant}
\en
In a rather more technical terms, one might choose to say that a one parameter translation group (a special subgroup of Arnold's symmetry group\cite{Arnold}) in the label space when represented by MHD load, MHD conjugate load and MHD metage has generated the conservation of cross helicity.
\section{Conclusion}
In short, we have introduced the concept of magnetohydrodynamic load in the case of MHD with untangled magnetic field lines and extended the definition for the non-trivial case of toroidal configuration of the magnetic field lines in MHD.
The elegant relation between the MHD loads and the magnetic helicity has been utilised to establish the relation (\ref{nl}) that very nicely connects the linking number of a pair of arbitrary magnetic field lines in the toroidal configuration discussed in the paper with the invariant loads in the magnetofluid having infinite conductivity.
The MHD analogue of Ertel's theorem is rederived for the advected scalar -- MHD load {\it i.e.}, it has been shown that the potential magnetic field w.r.t. magnetohydrodynamic load is conserved.
Moreover, the role of the MHD metage in generating the conservation of cross helicity in non-viscous non-resistive magnetofluid has been discussed.
\acknowledgements
One of the authors, SC, wants to thank Prof. J.K. Bhattacharjee for his constructive comments and discussions on an earlier draft of the paper.
Also, the financial support from CSIR (India), in the form of fellowship to the same author, is gratefully acknowledged.
Ayan's help in providing the authors with various relevant scholarly articles is heartily appreciated.

\end{document}